\documentclass[twocolumn,showpacs,preprintnumbers,amsmath,amssymb]{revtex4}
\usepackage{graphicx}% Include figure files
\usepackage{dcolumn}% Align table columns on decimal point
\usepackage{bm}% bold math
\usepackage{pstricks}
\begin{document}

\preprint{APS/123-QED}

\title{The exact probability distribution of saturating states in random
sequential adsorption}% Force line breaks with \\ 

\author{Masatomo Iwasa}
% \altaffiliation[Also at ]{Department of Physics, Nagoya University.}%Lines break automatically or can be forced with \\
\author{Kyohei Fukuda}%
\affiliation{%
Department of Physics, Nagoya University \\ 
 Nagoya, 464-8602, Japan \\
}%

%\affiliation{

\date{\today}% It is always \today, today,

\begin{abstract}
We consider the non-overlapping irreversible random sequential adsorption
 (RSA) process on one-dimensional finite line, which is known also
 as the car parking process.  
The probability of each coverage in saturating states is analytically and
 exactly obtained. 
In the derivation, a new representation of states in RSA process is
 introduced, which effectively works to make the calculation clear and
 simple. 
\end{abstract}

\maketitle

\section{Introduction}
We often see in various scientific and sociological fields that identical
objects are sequentially placed at random on a region
\cite{Evans1993} .  
Then, a meaningful question naturally arise: How much of the region is
finally covered by them? 
That kind of phenomenon is known as a non-overlapping irreversible random
sequential adsorption (RSA), and the theories have been mainly developed in
the field (see \cite{Evans1993} and references therein). 
It has been found that in arbitrary dimension, RSA process reach a
saturating or jamming configuration, where further adsorption events are
not possible. 
The final coverage as well as the temporal approach to the saturating
state are of interest
\cite{Evans1993,Feder1980,Widom1966,Pomeau1980,Swendsen1981,HinrichsenFederJossang1986,Renyi1958,GonzalezHemmerHoye1974}. 
Exact analytical results have been obtained mainly in one dimension,
where the problem is also known as the parking problem
\cite{Evans1993,Renyi1958,Widom1966,GonzalezHemmerHoye1974}.    

In those previous works, while the mean coverage has been obtained,
the probability for each coverage in saturating states has never been
revealed. 
Deriving it on one-dimensional RSA process analytically and exactly
is the main purpose of this paper.  

\section{The main result}
Suppose identical particles of unit length are placed sequentially
 at random on an interval of length $l$,
 subject to the constraint of no overlap. 
Here ``place particles at random'' means in this paper that each of them
 is placed based on the uniform distribution only on those empty spaces
 that have enough length to be placed. 
Note that it is not necessary to take the probability of finding
 already occupied place into account because we are not interested in
 the kinetics but only in the saturating states here. 
Then, the primary aim of this study is to obtain the probability of 
 saturating states where there are $n$ particles. 
The following is the main result, 
 and its derivation is accomplished through this paper. 
\\
\\
% {\bf The main result. }
 {\it In RSA process as above, 
 the probability of saturating states where there are $n$
 particles on the interval of length $l$ becomes
      \begin{eqnarray}
       P_{l,n}=   \left[
                   \prod_{r=0}^{n-1}
                   \left(1-Q_{l,r}\right)
                  \right]
                  Q_{l,n},\ 
       (0<n\leq l). 
\label{eq:2-1}
      \end{eqnarray}
Here 
      \begin{eqnarray}
       Q_{l,r}:=\left\{
                 \begin{array}{ll}
                  0,&r\leq \frac{l-1}{2},\\ 
		  \sum_{k=0}^{\lfloor l-r\rfloor}(-1)^{k}{}_{r+1}C_k 
                  \left(1-\frac{k}{l-r}\right)^{r},&\frac{l-1}{2}<r\leq l,
                 \end{array}
                \right. 
\label{eq:2-2}
      \end{eqnarray}
where ${}_nC_r:=\frac{n!}{r!(n-r)!}$ and $\lfloor\cdots\rfloor$ denotes
the floor function. \\  
}

\section{The proof}
Let us introduce some notations which are used
throughout this paper.  
Suppose now there are $n$ particles on the interval of length $l$.
Then the state can be represented by a point in $\mathbb{R}^{n+1}$.
Namely, if we arrange the each length of $n+1$ empty spaces between
particles and between particles at the ends and the edges of the interval from
the left side to the other, all of the states can be represented by a
set of $n+1$ positive real number, say
\begin{eqnarray}
 (a_0,a_1,\ldots,a_n);\ 0\leq a_i,\ i=0,1,\ldots,n.
\label{eq:3-1}
\end{eqnarray}
An example of this notation is shown in Fig. 1.\\
\\
{}
\hspace*{1.1cm}$a_0$
\hspace{1.3cm}$a_1$
\hspace{1.2cm}$a_2$
\hspace{1.5cm}$a_3$

\psline{<->}(0.5,0)(1.2,0)
\psline{<->}(2.2,0)(3.0,0)
\psline{<->}(4.0,0)(4.6,0)
\psline{<->}(5.6,0)(7,0)
\\
\\

\psline(.5,0)(7,0)
\psline(.5,.3)(.5,-.3)
\psline(7,.3)(7,-.3)
\psline(1.2,1)(2.2,1)
\psline(1.2,1)(1.2,0)
\psline(2.2,1)(2.2,0)
\psline(3.0,1)(4.0,1)
\psline(3.0,1)(3.0,0)
\psline(4.0,1)(4.0,0)
\psline(4.6,1)(5.6,1)
\psline(4.6,1)(4.6,0)
\psline(5.6,1)(5.6,0)

\psline{<->}(.5,0)(7,0)
\\
\hspace*{4cm}l
\begin{eqnarray}
 \Longleftrightarrow: (a_0,a_1,a_2,a_3)
\nonumber
\label{eq:3-2}
\end{eqnarray}
\begin{center}
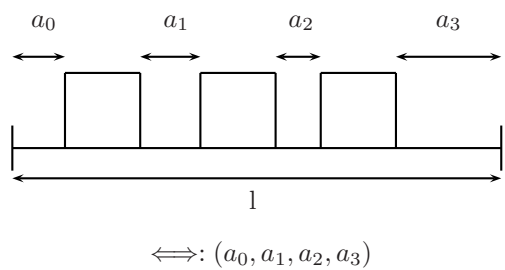
 Fig. 1. The representation of states: The state illustrated above
 where $n=3$ is expressed as $(a_0,a_1,a_2,a_3)\in \mathbb{R}^4$.\\  
\end{center}

Because the length of the interval is $l$, 
$a_i\ (i=0,1,\ldots,n)$ satisfies 
\begin{eqnarray}
 && \sum_{j=0}^n a_j = l-n.  
\label{eq:3-3}
\end{eqnarray}
Thanks to the constraint condition (\ref{eq:3-3}),
the number of independent valuables is reduced.
Henceforth in this paper we eliminate $a_0$.
Thus, any state where there are $n$ particles can be represented by a
point in $S_{n,l-n}$, 
\begin{eqnarray}
 (a_1,\ldots,a_n).
\label{eq:3-4}
\end{eqnarray}
Here $S_{n,p}\subset\mathbb{R}^{n}\ (0<p)$ is defined as 
\begin{eqnarray}
 S_{n,p}:=\left\{(a_1,\ldots,a_n)\Bigm| 
             \begin{array}{ll}
              0\leq a_i,& (i=1,\ldots,n),\\ 
              \sum_{j=1}^n a_j \leq p&
             \end{array}
            \right\}.
\label{eq:3-5}
\end{eqnarray}

In $S_{n,l-n}$, let us find the subset where all points correspond to
saturating states.  
Those are such states that any length of $n+1$ empty spaces are less than $1$.
Therefore, the subset corresponds to $R_{n,l-n}$.
Here $R_{n,p}\subset\mathbb{R}^n\ (0<p)$ is defined as 
\begin{eqnarray}
 R_{n,p}:=\left\{(a_1,\ldots,a_n)\Bigm| 
             \begin{array}{l}
              0\leq a_i<1,\ (i=1,\ldots,n)\\ 
              p-1<\sum_{j=1}^n a_j \leq p
             \end{array}
            \right\}.
\label{eq:3-6}
\end{eqnarray}
On the other hand, all points in $S_{n,l-n}\backslash R_{n,l-n}$
satisfy $1 \leq a_i $ for some $i=0,1,\ldots,n$.
That is to say, states corresponding to them are not saturating states,
and another particle can be placed on the interval. 

Next we discuss the probability of saturating states where there are $n$
particles on the interval of length $l$.
The derivation of the main result consists of three lemmas. 
At first, we see the following lemma.\\
{\bf Lemma 1.}
{\it Suppose the particles are placed at random on the interval.
Then, in terms of the notation $(\ref{eq:3-4})$,
each state represented by a point in $S_{n,l-n}$ has the same the
probability. }\\ 
{\bf Proof.}
The proof is accomplished by mathematical induction.
When $n=1$, the event is placing a particle on the interval which is
vacant.
Therefore, the claim is obviously true by means of the
definition of ``at random'', which is definitely described before the
statement of the main result.
Assume the lemma is true when $n=k$ for $k=1,2\ldots$.
Then we can show the lemma is also true when $n=k+1$ as below.
Any state represented by a point in $S_{k+1,l-(k+1)}$, say
$(a_1,\ldots,a_{k+1})$ according to the notation (\ref{eq:3-4}), occurs
if and only if a particle is placed 
in one of the $k+1$ different states represented by
$(a_2,a_3,\ldots,a_{k+1}),\  (a_1+a_2+1,a_3,\ldots,a_{k+1}),\
(a_1,a_2+a_3+1,\ldots,a_{k+1}),\ \cdots,\
(a_1,a_2,\ldots,a_k+a_{k+1}+1)$,  
all of which are points in $S_{k,l-k}$.
For example, a state $(3/2,2/5,5/4)$ occurs if and only if a particle is
placed in one of $(2/5, 5/4)$, $(19/10, 5/4)$ or $(3/2, 33/20)$.  
In short, any point in $S_{k+1,l-(k+1)}$ is mapped from $k+1$ different
points in $S_{k,l-k}$ by placing a particle. 
From the assumption of the lemma, these $k+1$ mapping have the same
probability. 
Therefore, it has been proved that each state represented by a point in
$S_{k+1,l-(k+1)}$ has the same probability if so does each state in
$S_{k,l-k}$. 
Thus, it has been inductively proved the lemma holds for all $n$.\ $\Box$ 

Now we can calculate the probability of saturating states
where there are $n$ particles.
In what follows, $V(A)$ denotes the volume of domain $A$. 
Suppose there are $n$ particles on the interval.
Then, from Lemma 1, the conditional probability that this state is
a saturating state in this given situation becomes 
\begin{eqnarray}
  \frac{V(R_{n,l-n})}{V(S_{n,l-n})}.
\label{eq:3-9}
\end{eqnarray}
On the other hand, the conditional probability that another
particle can be added to this state becomes
\begin{eqnarray}
  \frac{V(S_{n,l-n}\backslash
	    R_{n,l-n})}{V(S_{n,l-n})}.
\label{eq:3-10}
\end{eqnarray}
Therefore,
the probability of saturating states where there are $n$ particles on
the interval of length $l$, $P_{l,n}$, becomes,
\begin{eqnarray}
 P_{l,n}&=&\left[\prod_{r=0}^{n-1}
                 \frac{V(S_{r,l-r}\backslash R_{r,l-r})}
                      {V(S_{r,l-r})}
            \right]
                 \frac{V(R_{n,l-n})}{V(S_{n,l-n})},
\nonumber  \\
        &=&\left[\prod_{r=0}^{n-1}
                  \left(1-\frac{V(R_{r,l-r})}{V(S_{r,l-r})}\right)
            \right]
                  \frac{V(R_{n,l-n})}{V(S_{n,l-n})}.
\label{eq:3-11}
\end{eqnarray}

What should be done to complete the derivation of the main result is to
prove $\frac{V(R_{n,l-r})}{V(S_{n,l-r})}$ is equal to $Q_{l,r}$ in
Eq. (\ref{eq:2-2}). 
Then, let us calculate $V(S_{n,p})$ and $V(R_{n,p})$.
These are shown in Lemma 2 and Lemma 3.\\
{\bf Lemma 2.}
{\it 
\begin{eqnarray}
 V(S_{n,p})=\frac{p^n}{n!}.
\label{eq:3-12}
\end{eqnarray}
}
{\bf Proof.}
The domain $S_{n,p}$ reads
\begin{eqnarray}
 S_{n,p}=\left\{(a_1,\ldots,a_n)\Bigm| 
             \begin{array}{l}
              0\leq a_i\leq p-(a_1+\cdots+a_{i-1}),\ \\ 
              \hspace{0.5cm}(i=2,\ldots,n), \\
              0\leq a_1\leq p
             \end{array}
            \right\}.
\label{eq:3-13}
\end{eqnarray}
Therefore, the volume of $S_{l,p}$ becomes
\begin{eqnarray}
 V(S_{n,p})&=&\int_{0}^{p}da_1\int_{0}^{p-a_1}da_2
              \cdots\int_{0}^{p-(a_1+\cdots+a_{n-1})}da_n,
\nonumber \\
           &=&\frac{p^n}{n!}.\ \Box          
\label{eq:3-14}
\end{eqnarray}
By virtue of Lemma 2, let us find the volume of $R_{n,p}$.\\
{\bf Lemma 3.}
{\it
\begin{eqnarray}
 V(R_{n,p})=\left\{
              \begin{array}{ll}
	       0,& (n\leq p-1),\\
               \sum_{k=0}^{\lfloor
	       p\rfloor}(-1)^k{}_{n+1}C_k\frac{(p-k)^n}{n!},
                 & (p-1<n). 
	      \end{array}
             \right.
\label{eq:3-15}
\end{eqnarray}
}
{\bf Proof.}
First of all, 
let us obtain the volume of 
\begin{eqnarray}
 T_{n,p}:=\left\{(a_1,\ldots,a_n)\Bigm|
              \begin{array}{l}
	       0\leq a_i<1\ (i=1,\ldots,n)\\
               \sum_{j=1}^{n}a_j\leq p
	      \end{array} 
          \right\}.
\label{eq:3-16}
\end{eqnarray}\\
Once we obtain $V(T_{n,p})$,
$V(R_{n,p})$ is calculated from $V(R_{n,p})=V(T_{n,p})-V(T_{n,p-1})$.
To obtain the volume of $T_{n,p}$, we subtract the superfluous domain
from $S_{n,p}$. 
In this way, it follows
\begin{eqnarray}
 V(T_{n,p})&=&V(S_{n,p})
              -\sum_{1\leq i_1\leq n}V(U_{n,p}^{i_1})
              +\sum_{1\leq i_1<i_2\leq n}V(U_{n,p}^{i_1i_2})
 \nonumber \\
            &&+\cdots
              +(-1)^{k}\sum_{1\leq i_1<\ldots <i_{k}\leq n}
               V(U_{n,p}^{i_1\ldots i_k})
 \nonumber \\
            &&+\cdots
              +(-1)^{\lfloor p\rfloor}\sum_{1\leq i_1<\ldots
	      <i_{\lfloor p\rfloor}\leq n} 
               V(U_{n,p}^{i_1\ldots i_{\lfloor p\rfloor}}).
 \nonumber \\
\label{eq:3-17}
\end{eqnarray}
Here we have introduced for $k=1,\ldots,\lfloor p\rfloor$,
\begin{eqnarray}
 U_{n,p}^{i_1i_2\ldots i_k}
  &:=&\left\{(a_1,\ldots,a_n)\Biggm| 
             \begin{array}{l}
              1\leq a_i (i=i_1,\ldots,i_k),\\
              0\leq a_i (i\neq i_1,\ldots,i_k),\\ 
              \sum_{j=1}^{n}a_j\leq p\\
             \end{array}
            \right\}.
\label{eq:3-18}
\end{eqnarray}
Note that the formula (\ref{eq:3-17}) is suitable only for $\lfloor
p\rfloor\leq n$, that is $p\leq n+1$.
This is understandable if we concretely consider for n=3 as a simple
example.      
For $n+1<p$, from the definition of $T_{n,p}$ (\ref{eq:3-16}), it
obviously follows 
\begin{eqnarray}
 V(T_{n,p})=1.
\label{eq:3-17.5}
\end{eqnarray}
Let us calculate the right hand side of Eq. (\ref{eq:3-17}).
For any combination of $(i_1,\ldots,i_k)$,
\begin{eqnarray}
     V(U_{n,p}^{i_1i_2\ldots i_k})
      =V(U_{n,p}^{12\ldots k}),
\label{eq:3-19}
\end{eqnarray}
because of the symmetry with respect to the permutation of upper indices. 
Then, it follows
\begin{eqnarray}
      \sum_{1\leq i_1<\ldots<i_{k}\leq n}
     V(U_{n,p}^{i_1i_2\ldots i_k})
      ={}_nC_kV(U_{n,p}^{12\ldots k}).
\label{eq:3-20}
\end{eqnarray}
Therefore, Eq. (\ref{eq:3-17}) reads 
\begin{eqnarray}
 V(T_{n,p})&=&V(S_{n,p})
              +\left(\sum_{k=1}^{\lfloor p\rfloor }(-1)^{k}{}_nC_k
               V(U_{n,p}^{12\ldots k})\right).
\label{eq:3-21}
\end{eqnarray}
Let us obtain the volume of $U_{n,p}^{1\ldots k}$.
With the transformation
\begin{eqnarray}
 \tilde{a_i}:=\left\{
\begin{array}{ll}
 a_i-1, &(i=1,\ldots,k), \\
 a_i,   &(i=k+1,\ldots,n), 
\end{array}
 \right.
\label{eq:3-22}
\end{eqnarray}
$U_{n,p}^{1,\ldots,k}$ is written by
\begin{eqnarray}
  \left\{(\tilde{a_1},\ldots,\tilde{a_n})\Biggm| 
             \begin{array}{l}
              0\leq\tilde{a_i} (i=i_1,\ldots,n)\\
              \sum_{j=1}^{n}\tilde{a_j}\leq p-k\\
             \end{array}
            \right\},
\label{eq:3-23}
\end{eqnarray}
which corresponds to $S_{n,p-k}$.
Therefore, since the Jacobian of the transformation is one,
we obtain with the aid of Lemma 2
\begin{eqnarray}
 V(U_{n,p}^{1\ldots k})&=&V(S_{n,p-k}),\\
                       &=&\frac{(p-k)^n}{n!}.
\label{eq:3-24}
\end{eqnarray}
By substituting Eq. (\ref{eq:3-14}) and Eq. (\ref{eq:3-24}) into
Eq. (\ref{eq:3-21}), and according to Eq. (\ref{eq:3-17.5}),  we obtain 
\begin{eqnarray}
 V(T_{n,p})=\left\{
             \begin{array}{ll}
	      1,&(n\leq p-1), \\
              \sum_{k=0}^{\lfloor
             p\rfloor}(-1)^{k}{}_nC_k\frac{(p-k)^n}{n!},&(p-1<n).
	     \end{array}\ 
            \right. 
\label{eq:3-25}
\end{eqnarray}
Hence, we can immediately show that Eq. (\ref{eq:3-15}) holds
according to 
$V(R_{n,p})=V(T_{n,p})-V(T_{n,p-1})$,
noting that $V(T_{n,p})=1$ for $\lfloor p\rfloor=n$ while
Eq. (\ref{eq:3-17}) holds.\ $\Box$  

As noted before, what must be done to complete the derivation of the
main result is to show
that $\frac{V(R_{r,l-r})}{V(S_{r,l-r})}$ is equal to $Q_{l,r}$ in
Eq. (\ref{eq:2-2}).
By substituting Eq. (\ref{eq:3-12}) and Eq. (\ref{eq:3-15}),
We can immediately show that  
\begin{eqnarray}
 \frac{V(R_{n,p})}{V(S_{n,p})}
  &=&\left\{
      \begin{array}{ll}
       0,& (n\leq p-1), \\
       \sum_{k=0}^{\lfloor p\rfloor}(-1)^k{}_{n+1}C_k
       \left(1-\frac{k}{p}\right)^n, & (p-1< n).
      \end{array}
     \right. 
\label{eq:3-26}
\end{eqnarray} 

\section{Concluding Remarks}
The RSA model on one-dimensional finite interval has been analyzed exactly.  
As a result, the probability of each coverage at saturating states has
been obtained explicitly for a generic system.  
Then we can obtain the explicit expression of all moments.
The main result, that is exact probability for any system size, would be
useful for explaining the various actual RSA phenomena since experiments
are always carried out on a system of finite size.  
Compared with the previous studies
\cite{Evans1993,Feder1980,Widom1966,Pomeau1980,Swendsen1981,HinrichsenFederJossang1986,Renyi1958,GonzalezHemmerHoye1974},
this study takes another way to represent a state during the adsorbing
process. 
Through the derivation of the main result, we see that this notation
effectively works to make the calculation clear and simple. 
Besides this, omitting the kinetics also seems to be an important point 
to obtain the result.
However, since we are almost always interested in not only final states
but also kinetics in physics, construction of an appropriate master equation
utilizing this result seems to be an significant problem for the future.   
The consistency with the previous studies
\cite{Evans1993,Feder1980,Widom1966,Pomeau1980,Swendsen1981,HinrichsenFederJossang1986,Renyi1958,GonzalezHemmerHoye1974}
is another question.   
This should be investigated, if possible, by obtaining the
asymptotic behavior of the mean coverage, that is the average of $n/l$,
for $l\rightarrow\infty$.

\end{document}